\documentclass[sigconf]{acmart}
\usepackage{amsmath}
\usepackage{amsfonts}
\usepackage[utf8]{inputenc}
\usepackage[ruled, vlined]{algorithm2e}
\usepackage{amsopn}
\usepackage{subcaption}
\usepackage{hyperref}
\usepackage{cleveref}
\crefname{app}{appendix}{Appendix}
\crefname{ineq}{Inequality}{Inequalities}
\creflabelformat{ineq}{#2{\upshape(#1)}#3}

\DeclareMathOperator{\Var}{Var}

\newcommand{\pluseq}{\mathrel{+}=}

\setcopyright{acmlicensed}
\copyrightyear{2026}
\acmYear{2026}
\acmConference[Conference acronym 'XX]{Make sure to enter the correct
  conference title from your rights confirmation email}{June 03--05,
  2018}{Woodstock, NY}
\begin{document}

\title{Walking on Spheres and Talking to Neighbors: Variance Reduction for Laplace's Equation}

\author{Michael T. Czekanski}
\email{mc2589@cornell.edu}
\affiliation{
  \institution{Cornell University}
  \city{Ithaca}
  \state{NY}
  \country{USA}
}

\author{Benjamin J. Faber}
\affiliation{
  \institution{University of Wisconsin - Madison}
  \city{Madison}
  \state{WI}
  \country{USA}
}

\author{Margaret E. Fairborn}
\affiliation{
  \institution{Columbia University}
  \city{New York}
  \state{NY}
  \country{USA}
}

\author{Adelle M. Wright}
\affiliation{
  \institution{University of Wisconsin - Madison}
  \city{Madison}
  \state{WI}
  \country{USA}
}

\author{David S. Bindel}
\affiliation{
  \institution{Cornell University}
  \city{Ithaca}
  \state{NY}
  \country{USA}
}

\renewcommand{\shortauthors}{Czekanski et al.}

\begin{abstract}
Walk on Spheres algorithms leverage properties of Brownian Motion to create Monte Carlo estimates of solutions to elliptic partial differential equations. 
We propose a new caching strategy that leverages the continuity of paths of Brownian Motion.
Until recently, estimates were constructed pointwise and did not use the relationship between solutions at nearby points within a domain. 
In the case of Laplace's equation with Dirichlet boundary conditions, our algorithm has improved asymptotic runtime compared to previous approaches.
Our results are achieved by information reuse from a cache of fixed size.
We also provide bounds on the performance of our algorithm and demonstrate our approach on example problems of increasing complexity. 
\end{abstract}



\begin{teaserfigure}
    \centering
    \begin{subfigure}[b]{0.3\textwidth}
        \includegraphics[width=\textwidth]{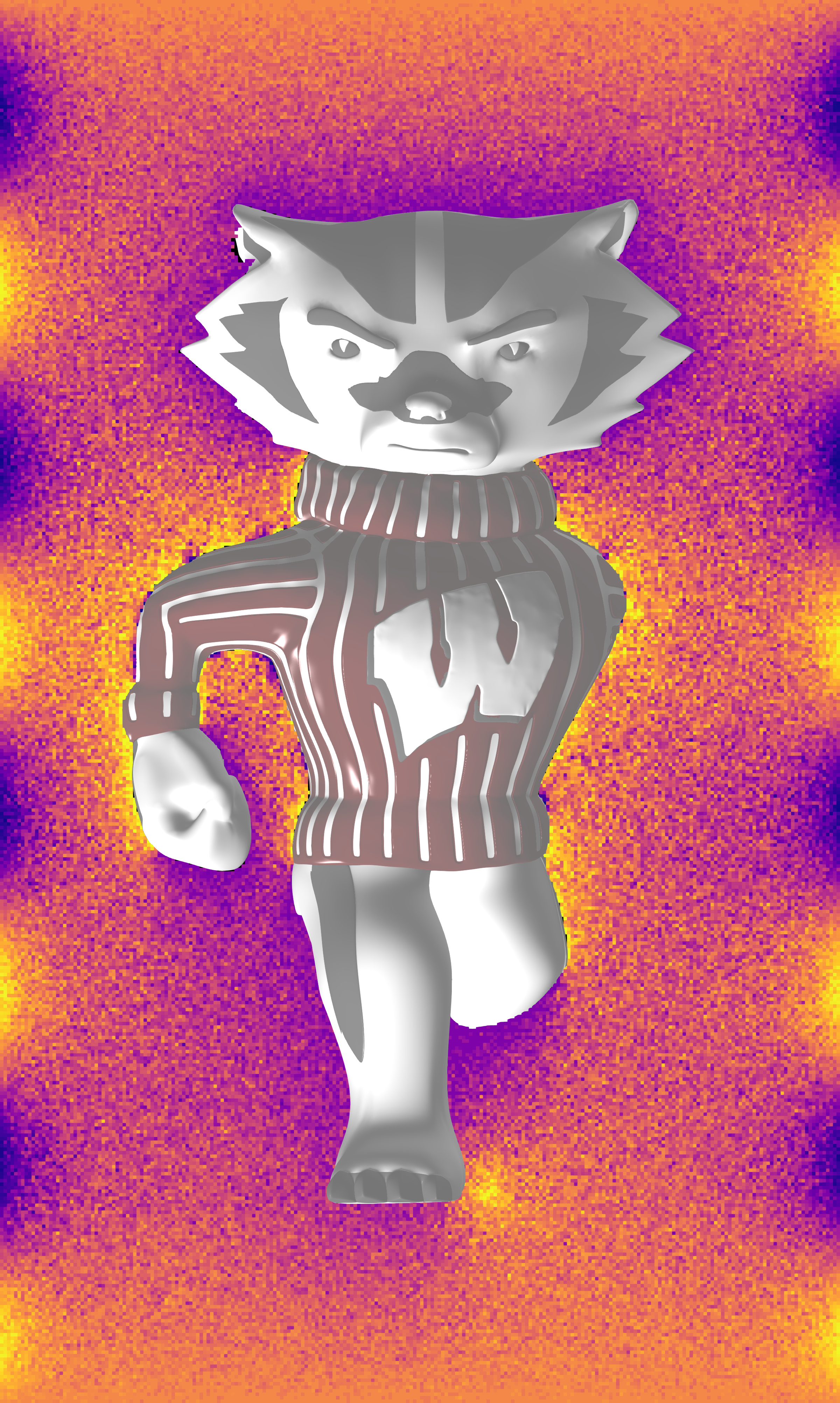}
        \caption{Standard Walk on Spheres}
    \end{subfigure}
    \begin{subfigure}[b]{0.3\textwidth}
        \includegraphics[width=\textwidth]{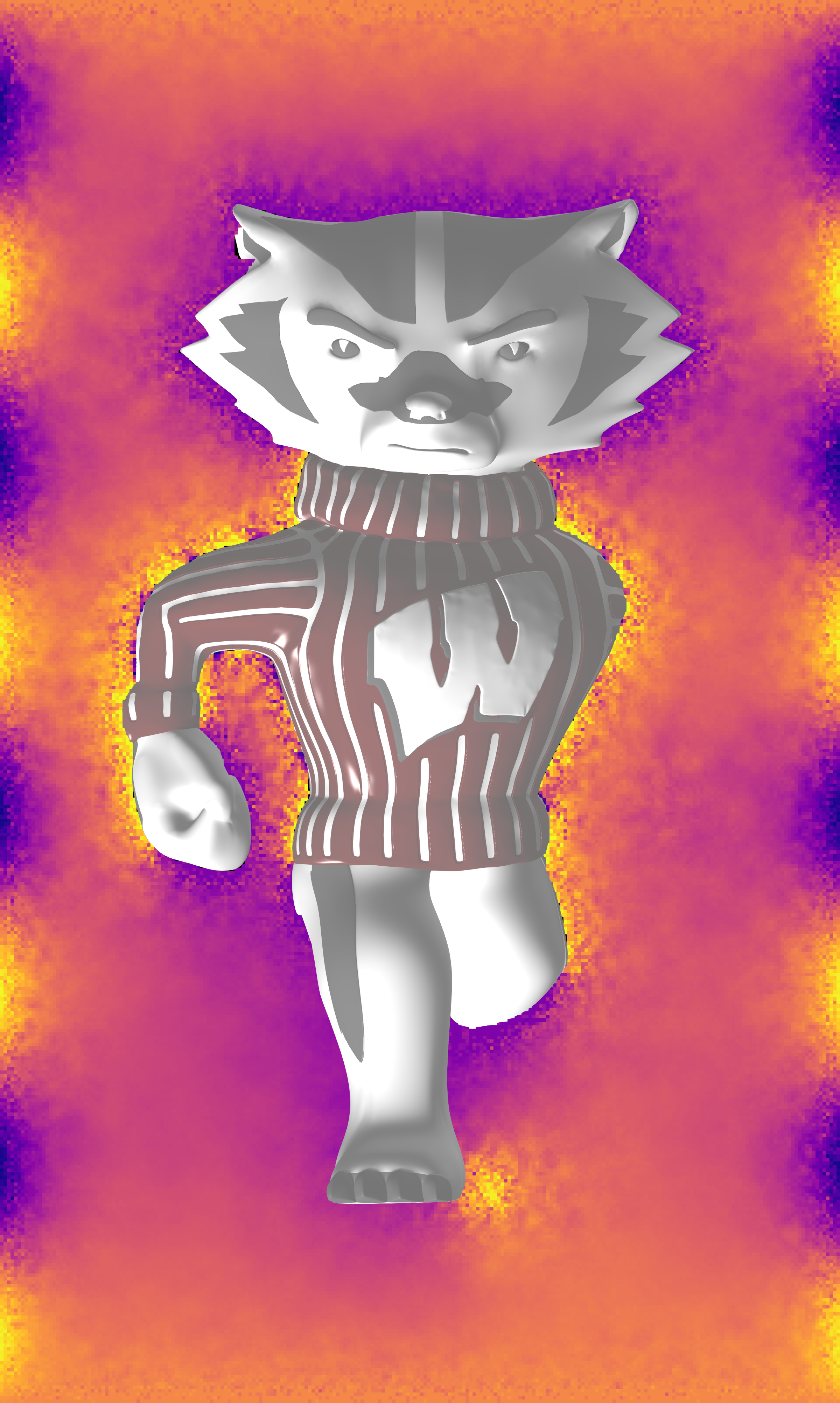}
        \caption{Equal Weighted Info. Reuse (ours)}
    \end{subfigure}
    \begin{subfigure}[b]{0.3\textwidth}
        \includegraphics[width=\textwidth]{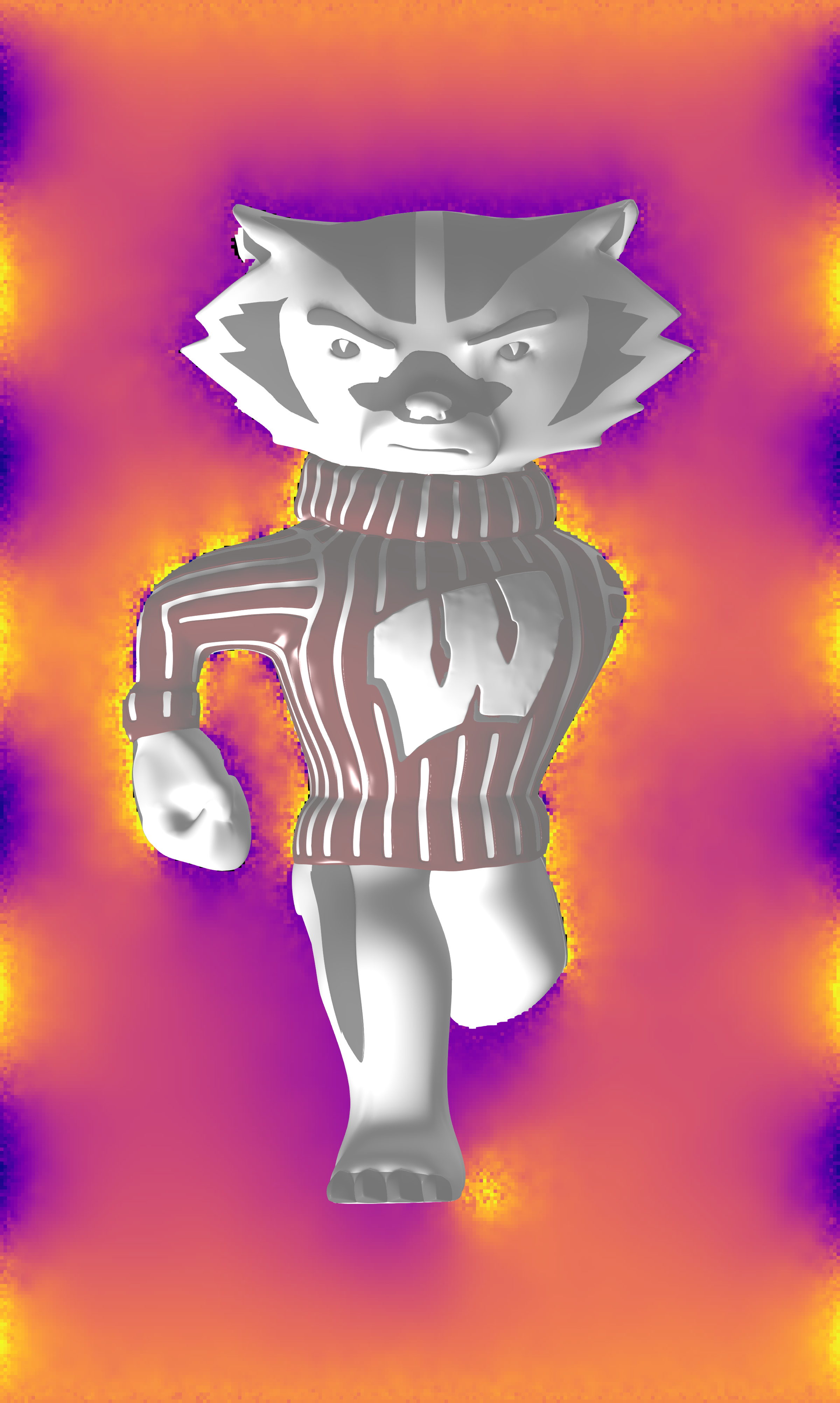}
        \caption{Variance Weighted Info. Reuse (ours)}
    \end{subfigure}
    \caption{Standard Walk on Spheres output (left) vs. our variance equal-weighted (middle) and inverse variance-weighted (right) reuse strategies with 10 walks per pixel. Time for information reuse is negligible (< 1$\%$ of time to perform walks)} \label{fig:teaser}
\end{teaserfigure}

\maketitle

\section{Introduction}
Physics-based modeling plays an important role in enhancing the realism of virtual environments. 
We do this by solving ordinary or partial differential equations (PDEs).
A particularly important class of PDEs, often arising in the description of steady-state systems, are elliptic PDEs. 
The most fundamental of these, Laplace's equation, describes situations arising naturally in heat conduction, fluid dynamics, and electrostatics.

Often we need to obtain solutions to Laplace's equation on geometrically complex domain representing real-world objects and environments.
Generally, this must be done numerically.
There exist a variety of mesh-based and mesh-free methods each with distinct advantages.
We present improvements to the mesh-free Walk on Spheres algorithm which has two advantages: it is not necessary to discretize the domain, and solutions can be computed on subsets of the domain without the need to compute the global solution.

\subsection{Laplace's Equation}
We consider 
Laplace's equation 
on a regular, connected domain $\Omega\subset\mathbb{R}^d$, with Dirichlet conditions on the boundary $\partial \Omega$:
\begin{align}
\begin{cases} \label{eq:laplace}
\Delta u(x) = 0 & x \in \Omega \\
u(x) = f(x) & x \in \partial \Omega
\end{cases}
\end{align}
where $\Delta = \nabla \cdot \nabla$, $x\in\mathbb{R}^d$ and $f$ is continuous and bounded.

In simple domains such as spheres and polyhedra, analytic solutions can be found via separation of variables and Green's function methods.
For more complex domains, many different numerical methods exist to solve Eq. (\ref{eq:laplace}). 
Broadly, we categorize these into mesh-based and mesh-free methods.

Mesh-based methods, such as finite-difference, finite-element, and finite-volume methods, rely on spatial discretizations of the domain on which the solution is approximated.
Complicated domains with rough boundaries or sharp points can require a very fine discretization to accurately represent the boundary and resolve strong gradients in the solution.
These approximations are computed as the solution to a linear system with size depending on the granularity of the discretization. 
In applications such as thermal analysis of a mechanical system, we may only be interested in the solution near a critical component.
In this case, it is inefficient to globally solve Laplace's equation when we are only interested in the solution on a small subset of the domain.

Mesh-free methods are particularly well-suited for problems on domains with complex geometry. 
Techniques include radial basis function methods, boundary integral equations and the method of particular solutions.
However, as with mesh-based methods, obtaining solutions using these methods requires solving a global system of equations.

\subsection{Contribution}
In this work, we present improvements to the Walk on Spheres algorithm \cite{muller1956some}, 
a mesh-free Monte Carlo method for solving Laplace's equation.
Walk on Spheres generates Monte Carlo samples by following random walks from the point of interest until they reach the boundary.
It requires neither a discretization of the domain nor a global solution.

We develop a novel variance reduction method for Walk on Spheres based on information reuse and demonstrate its performance on several example problems in two and three dimensions.
By exploiting the (almost sure) continuity of paths of Brownian Motion, we enable information reuse from walks started at nearby points to reduce variance.
We then use these methods to enable a caching scheme that reduces the required number of samples.
The results we present here can be generalized to other settings where Walk on Spheres is used, including Poisson equations, equations with spatially varying coefficients, and problems with Neumann boundary conditions \cite{sawhney2020monte, sawhney2022grid}. 
Additionally, we provide bounds on the variance of our estimators and gesture towards generalization to other classes of PDEs.

\section{Walk on Spheres Algorithm} \label{sec:wos}

The original Walk on Spheres algorithm  is presented in \Cref{alg:wos}.
It exploits the connection between elliptic partial differential equations and Brownian Motion.

\begin{theorem}\label{thm:stochastic_sol}
    \cite{kakutani1944143, muller1956some}
    Given a sufficiently regular domain $\Omega$ and a continuous function $f$ on $\partial \Omega$, the solution to \Cref{eq:laplace} is 
    \begin{align}
    u(x) &= \mathbb{E}_x[f(W_\tau)] 
    \end{align}
    where $\tau = \inf\{t > 0 : W_t \notin \Omega \}$, $W_t$ is a path of Brownian Motion, and $\mathbb{E}_x$ is the expectation under the measure implied by conditioning on $W_0 = x$.
    Note that $\tau$ is random under realizations of paths of Brownian Motion. It is the first time that the path leaves the domain $\Omega$.
\end{theorem}

Walk on Spheres traces a path of Brownian Motion by only observing its location when it first exits a (random) sequence of spheres (although other shapes, such as rectangles, can be used \cite{deaconu2006random}).

\begin{algorithm}
    \caption{Original Walk on Spheres for Laplace's Equation \cite{muller1956some}} \label{alg:wos}
    \SetKwInOut{Input}{Input}\
    \DontPrintSemicolon
    \Input{Points of interest $x_1, \ldots, x_m \in \Omega$,  continuous $f$ on $\partial \Omega$, number of walks to perform $n$, $\varepsilon > 0$}
    \For{$i \in [m]$}{
        $\hat{u}(x_i) = 0$ \;
        \For{$j \in [n]$}{
            $\text{loc} = x_i$ \;
            $r = \inf_{y \in \partial \Omega} \operatorname{dist}(loc,y)$ \;
            \While{$r> \varepsilon$}{
            loc = uniform sample from $\partial B(loc, r)$ \; 
            $r = \inf_{y \in \partial \Omega} \operatorname{dist}(loc,y)$ 
            } 
            $\hat{u}(x_i) \pluseq f(\operatorname{argmin}_{y \in \partial \Omega} \operatorname{dist}(loc,y) )$
        }
    }
    $\hat{u}(x_i) = \frac{1}{n} \hat{u}(x_i)$ for all $i$
  \end{algorithm}

\Cref{alg:wos} terminates walks when they are within $\varepsilon$ of the boundary, while \Cref{thm:stochastic_sol} requires that the walk exactly reach the boundary.
It would be impractical to let $\varepsilon = 0$, as at each step the walk would generally terminate with probability zero.
Instead we let $\varepsilon >0$ to ensure a positive probability of termination at each step.
This incurs an $O(\varepsilon)$ bias in our estimates, but this can be improved using multi-fidelity methods over choices of $\varepsilon$ \cite{pauli_multilevel_2014}.
The runtime also depends on $\varepsilon$, as small values require longer walks to the boundary.
The number of steps in a single walk is $O\left(\log \varepsilon^{-1}\right)$ \cite{binder2012rate} and the runtime of \Cref{alg:wos} is $O(mn\log \varepsilon^{-1})$.

While tangent to the information reuse scheme here, the dominant computational cost of observing a walk on spheres is the boundary distance calculation.
The details of this computation are not present in the Walk on Spheres algorithm, but in practice we use a bounding volume hierarchy as implemented in ImplicitBVH.jl \cite{Chitalu2020implicitbvh}.
This enables parallelized computation of boundary distances across batches of walks.

\subsection{Related Work}

Solutions obtained using Walk on Spheres have a salt-and-pepper appearance as a result of the independent statistical noise in our estimators.
The most straightforward way to reduce the variance is to increase the number of walks $n$, which is both expensive and embarassingly parallel.
Other approaches have altered walks to reduce variance \cite{bakbouk2023mean, milstein1994numerical, zou2004robust}, altered walks to reduce runtime \cite{hwang2015off}, or leveraged a smaller number of walks more effectively \cite{miller2023boundary, sawhney2020monte,qi2022bidirectional,booth1982regional}.

Our approach falls into this last category by reusing information from walks started at other points in $\Omega$.
However, it is distinguished from previous works by the fact that (i) we allow the starting point of the walks to be chosen arbitrarily as opposed to sampled randomly \cite{bakbouk2023mean,miller2023boundary} and (ii) we also provide bounds on the variance of our estimators.
Our strategy uses the spatial relationships of $u$ to reduce the variance of the Walk on Spheres estimator through information reuse.
Other caching schemes have applied similar ideas to broader classes of PDEs in concurrent work \cite{bao2025off,zhou2025harmonic}.
Both rely on the off-centered Poisson kernel, but leverage
it in different ways (see Supplemental Material of Zhou for more).
Bao et al. \shortcite{bao2025off} uses bounds for Laplace's equation in more general settings without modification.
Zhou et al. \shortcite{zhou2025harmonic} cache estimates of Fourier coefficients while we instead cache the results of walks. 
Our work remains the only information reuse scheme for Walk on Spheres with analytic guarantees.

Our information reuse algorithm, outlined in \Cref{sec:info_reuse}, is implemented as post-processing of walks from the original Walk on Spheres.
We provide a condition on $\|x_i - x_j\|$ where termination points of walks started at $x_j$ can be used to estimate $u(x_i)$, increasing the number of samples used to estimate $u(x_i)$, such that the variance is reduced.
In \Cref{sec:var_reduce}, we propose two information reuse algorithms; in \Cref{sec:caches}, we show that they only require walks to start on a grid; and in \Cref{sec:experiments}, we show the performance on several example problems.

\section{Information Reuse}  \label{sec:info_reuse}

Walk on Spheres observes exit points for a path of Brownian Motion along a sequence of spheres.
This guarantees that the path has not yet left the domain because the path must leave the current sphere before leaving the domain.
A similar argument holds for off-centered spheres.
For a given sphere contained in our domain, walks started from two points within the sphere must both exit the sphere before reaching the boundary.
In fact, for paths started from any two points within the sphere, the exit distributions are known and mutually absolutely continuous.
This allows us to reuse walks as though they had been importance sampled.
The relationship between these distributions on the unit disk is stated in \Cref{thm:ev_sphere}.

\begin{theorem}
    \cite{durrett2005probability}  \label{thm:ev_sphere}
    Let $\Omega = \{x : \|x\|_2 < 1 \} \subset \mathbb R^d$ and $\tau = \inf \{t : W_t \notin \Omega \}$. If $f$ is bounded and continuous, then
    \begin{align}
    \mathbb{E}_x \left[f(W_{\tau}) \right] &= \frac{1}{|\partial \Omega|}\int_{\partial \Omega} k(x, y) f(y) dy \\
    k(x, y) &= \frac{1 - \|x\|^2}{\|x - y\|^d}\label{eq:pois_kernel}
    \end{align} 
    is the solution to Laplace's equation. A deterministic treatment of the right-hand side as the solution to Laplace's equation can be found in Courant \cite{courant1962}. 
\end{theorem}

Before proving that \Cref{thm:ev_sphere} enables our information reuse scheme, we first generalize \Cref{thm:ev_sphere} to arbitrary spheres by mapping them to the unit sphere \cite{courant1962}.
On a sphere of radius 
$r$ centered at $c$, which we denote by $\partial B(c, r)$, we can use \Cref{thm:ev_sphere} replacing the Poisson kernel $k(x,y)$ on the unit sphere by
\begin{align}
    k_c^{(r)}(x, y) &= \frac{1 - r^{-2}\|x - c\|^2}{r^{-1} \|x - y \|^d}. \label{eq:gen_pois_kernel}
\end{align}
We will proceed with this more general definition. 
This produces estimates of $u(x)$ using walks started at $x'$, provided that both $x$ and $x'$ are in some sphere entirely within $\Omega$.

\begin{corollary} \label{corr:reweight}
    Consider a sphere $\partial B(c,r) \subseteq \Omega$. The solution to \Cref{eq:laplace} at any point $x \in B(c,r)$ can be computed from paths of Brownian Motion started at $c$ by a change of measure. Let $\tau$ be defined as before and let $\tau_B$ be the first time the path of Brownian Motion exits $B(c,r)$.  Then we have
    \begin{equation}
        \mathbb{E}_{x} \left[f(W_{\tau}) \right] = \mathbb{E}_{c} \left[k_c^{(r)}(x, W_{\tau_B})f(W_{\tau}) \right]. \label{eq:laplace_reweight}
    \end{equation}
\end{corollary}
\begin{proof}
    $B_\tau|B_0=c$ is uniformly distributed over $\partial B(c,r)$. By \Cref{thm:ev_sphere}, the two integrals are equivalent.
\end{proof}

Given points $x$ and $x'$ on the interior of $\Omega$ in \Cref{eq:laplace} such that $x' \in B(x,r) \subset \Omega$,
we can recover information about $u(x')$ from our estimates of $u(x)$ by reweighting walks based on their first step.
In post-processing walks, we have already observed samples of $W_\tau^c$, which can be weighted by $k_c^{(r)}(x, W_{\tau_B})$ to produce unbiased estimates of
$u(x')$.

The information-reuse estimator for a walk that begins at $x_i$ and is used to estimate $u(x)$ is
\begin{align} \label{eq:ir_est}
\hat u_{x_i}(x) = k_{x_i}(x, W_{\tau_B}^{x_i}) f(W_\tau^{x_i}).
\end{align}
Note that we drop the superscript $r$ on the Poisson kernel $k$ for simplicity. When considering spheres centered at $x_i$ (or any other point), we
assume $r$ is as large as possible such that $B(x_i,r)$ is contained in $\Omega$.
This estimator re-weights the observed walks by the Poisson kernel to obtain an unbiased estimate at the new point $x$.

Similar reweighting was used by Hwang et al. \cite{hwang2015off} to reduce the average number of steps in a walk by using a larger off-centered sphere at the beginning of a walk.
They observe fewer steps per walk, but higher variance. This follows from \Cref{lem:var_tradeoff} because of the variance in the weighting term.
Walks started at other points have been used, but the starting points of those walks had to be randomly sampled \cite{bakbouk2023mean, miller2023boundary}.
In contrast, \Cref{corr:reweight} provides a condition where walks started at $c$ can be used to estimate $u(x)$ regardless of how $c$ is chosen.

\section{Variance Bounds} \label{sec:var_reduce}

Having constructed our information reuse estimators, we now compare their variance to the original Walk on Spheres.
For a given point $x \in \Omega$, the variance of the original Walk on Spheres estimator can be bounded by Popoviciu's inequality
\begin{equation}
    \Var(\hat u(x)) = \mathbb{E}\left[\hat u (x)^2\right] - \mathbb{E}\left[ \hat u(x) \right]^2 \le \frac{1}{4}M^2 
\end{equation}
where $M = \max_{x \in \partial \Omega} f(x)$ and we assume $\min_{x \in \partial \Omega}f(x)=0$ without loss of generality.
The original Walk on Spheres estimator with $n$ walks then has worst-case variance bounded by $n^{-1}M^2/4 = O(n^{-1})$.
For estimators of $u(x)$ from walks that began at locations other than $x$, we require that this bound still holds.
Otherwise, there are potentially choices of $f$ and $\Omega$ where our final estimators perform worse than the original Walk on Spheres algorithm.

The variance of the information reuse estimator (\Cref{eq:ir_est}) can be bounded using deterministic bounds on the Poisson kernel and boundary function.

\begin{lemma} \label{lem:var_tradeoff}
Let $x, x' \in \Omega \subset \mathbb{R}^d$ such that $x' \in B(x,r) \subseteq \Omega$.
\begin{align}
    \Var \left(\hat u_x(x') \right) &\le 
    \frac{1}{4} 
    \frac{\left(1 - r^{-2} \|x-x'\|^2 \right)^2}{\left(1 - r^{-1}\|x-x'\|\right)^{2d}} M^2
\end{align}
\end{lemma}

\begin{proof}
The Poisson kernel is bounded above:
\begin{align}
 k(x, y) &\le \frac{1 - r^{-2}\|x - x'\|^2}{(1 - r^{-1}\|x - x'\|)^{d}}
\end{align}
Use this to bound the variance again by Popoviciu's inequality:
\begin{align}    
    \Var \left(\hat u_x(x') \right) &=  \Var \left( k_x(x', W_{\tau_1}^{x}) f(W_\tau)\right) \\
    &\le  \mathbb{E}_x[k_x(x', W_{\tau_1}^{x})^2 f(W_\tau^x)^2] \\
    &\le  \frac{1}{4}\frac{\left(1 - r^{-2}\|x - x'\|^2\right)^2}{(1 - r^{-1}\|x - x'\|)^{2d}} M^2.
\end{align}
\end{proof}

As $\|x - x'\|$ approaches $r$, the variance of $\hat u(x)$ can increase rapidly. 
Therefore, passing information from all nearby points without accounting for this potentially increased variance can produce an estimator with larger variance than the original estimator, as observed in experiments by Hwang et al. \cite{hwang2015off}.

This bound holds for Laplace's equation and similar bounds can be derived for the Poisson equation and other elliptic PDEs.
In general, solutions to PDEs viewed as expectations over paths of Brownian Motion can partitioned into the first step and later steps.
Results from later steps can be re-weighted in this way, while the first step would need to be re-estimated. In the case of Laplace's equation the only contribution comes form the terminal location, allowing us to focus on $f$ at the boundary.

\subsection{Equal Weighting} \label{sec:eql_wgt}

We begin by considering information reuse using a hard cutoff and equal-weighting the available estimators.
We present this approach in \Cref{alg:eql} where $x_{l,j}^1$ and $x_{l,j}^*$ are respectively the first step
and final location of walk $j$ from $x_l$.

For clarity, we present \Cref{alg:eql} as nested for loops, but in practice we efficiently find pairs $x_i, x_l$ where information can be passed using a parallelized bounding volume hierarchy and then parallelize over the walks that can be shared.
For example, if solving on a 256x256 grid with 10 walks per point, the entirety of our cache can be stored in less than 6MB and we only need to check $256^2$ pairs to see if information can be shared.
All of this can be performed in parallel  using traversals of a BVH. We later show that the runtime impact is minimal.
The data for this operation is reduced by batching walks and performing variance reduction for each batch on the fly.

\begin{algorithm}
    \caption{Equal Weighted Algorithm} \label{alg:eql}
    \SetKwInOut{Input}{Input}
    \DontPrintSemicolon
    \Input{Walk data from \Cref{alg:wos}}
    \For{$i \in [m]$}{
        \For{$l \in [m]$}{
            \If{$\|x_i - x_l\| < C(d) r_l$}{
                \For{$j \in [n]$}{
                    $\text{sums[i]} \pluseq k_{x_l}(x_i, x_{l,j}^1) f(x_{l,j}^*)$ \;
                    $\text{weights[i]} \pluseq 1$\;
                }
            }
        }
        $\hat{u}(x_i) = $ \text{sums[i] / weights[i]}  for all $i$\;
    }
\end{algorithm}

To establish a cutoff, we observe that we can use \Cref{lem:var_tradeoff} to obtain a constant $C(d)$ that restricts the variance of the information reuse estimator.
For each dimension $d$, there exists a constant $C(d) \in [0,1]$ such that  $r^{-1}\|x-x'\| \le C(d)$ implies $\Var \left(\hat u_x(x') \right) \le \frac{3}{4}M^2$, given by
\begin{align}
    \frac{(1+C(d))^2}{(1-C(d))^{2d-1}} \le 3.
\end{align}
Existence is guaranteed because the left-hand side is continuous, 1 when $C(d)=0$, and tends to $\infty$ when $C(d)\to 1$.
We observe that $C(2) \le \frac{\sqrt{3}-1}{\sqrt{3}+1} \approx0.268$, and $C(3) \le 0.176$.
Tightening these bounds leads to more availability of information reuse and subsequently lower variance.
In supplemental material, we improve on this bound to obtain $C(2) \le 0.447$.

When equally weighting our information reuse estimators  with walks from $p-1$ nearby points where $\|x_i -x_l\| \le  C(d) r_l$, we can write the final estimator as
\begin{align}
    \hat u(x_i) &= \frac{1}{np}\sum_{j=1}^n f(x_{i,j}^*) + \frac{1}{np}  \sum_{x_l : \|x_i - x_l\| < C(d) r_l}\sum_{j=1}^n k_{x_l}(x_i, x_{l,j}^1)f(x_{l,j}^*)
\end{align}
and bound its variance using \Cref{lem:var_tradeoff}
\begin{equation}
    \Var(\hat u(x_i)) \leq \frac{1}{4np^2} M^2 + \frac{3(p-1)}{4np^2}M^2 = \frac{3p-2}{4np^2}M^2.
\end{equation}

This recovers the worst-case bound of the variance for the original Walk on Spheres algorithm, while demonstrating the benefit of information reuse as the variance is now $O(n^{-1}p^{-1})$.
Note that $n$ can be increased by running more walks and $p$ can be increased by choosing starting points more carefully, or obtaining a tighter bound than \Cref{lem:var_tradeoff}.

\subsection{Variance Weighting}

For a collection of unbiased estimators, the minimum variance  linear combination that is still unbiased, weights estimators inversely by their variance.
Given that the variance of our estimators is unknown, we substitute the variance bound of \Cref{lem:var_tradeoff}.
Instead of a hard cutoff to include estimators, this allows us to reuse any available information with more weight on walks started from closer points.

Using the variance bound from \Cref{lem:var_tradeoff}, this estimator is 
\begin{align}  \label{eq:inv_var_wgt}
\hat u(x) &= \left(\sum_{i=1}^p w_i\right)^{-1} \sum_{i=1}^p w_i \hat u_{x_i}(x) \\
\text{where } w_i &= \frac{(1 - r_i^{-1}\|x - x_i\|)^{2d}} {\left(1 - r_i^{-2}\|x - x_i\|^2\right)^2}
\end{align}
and $r_i$ is the radius of $B(x_i)$. 
We can again bound the variance of this estimator.
\begin{align}
\Var \left(\left(\sum_{i=1}^p w_i\right)^{-1} \sum_{i=1}^p w_i \hat u_{x_i}(x) \right) &= \left(\sum_{i=1}^p w_i\right)^{-2} \sum_{i=1}^p w_i^2 \Var \left(\hat u_{x_i}(x) \right) \\
&\le \frac{1}{4}\left(\sum_{i=1}^p w_i\right)^{-2} \sum_{i=1}^p w_i M^2 \\
&= \frac{1}{4}\left(\sum_{i=1}^p w_i\right)^{-1} M^2 \\
&< \frac{1}{4}\min_{i \in [p]} \frac{M^2}{w_i} \label[ineq]{ineq:loose_var_bd} \\
&= \frac{1}{4}\frac{M^2}{\max_{i \in [p]} w_i} \le \frac{1}{4}M^2.
\end{align}

Asymptotically, this bound is not as strong as the equal weighted estimator, because the weights are not bounded away from $0$ and 
Ineq. 22 is rather loose.
However, with no restriction on $\|x - x_i\|$, information can be passed from more points.
The computation of this estimator is presented in \Cref{alg:var}

\begin{algorithm}
    \caption{Variance Weighted Algorithm} \label{alg:var}
    \SetKwInOut{Input}{Input}
    \DontPrintSemicolon
    \Input{Walk data from \Cref{alg:wos}}
    \For{$i \in [m]$}{
        \For{$l \in [m]$}{
            \If{$\|x_i - x_l\| < r_l$}{
                $w = \frac{(1 - r_i^{-1}\|x_i - x_l\|)^{2d}} {\left(1 - r_i^{-2}\|x_i - x_l\|^2\right)^2}$ \;
                \For{$j \in [n]$}{
                    $\text{sums[i]} \pluseq w k_{x_l}(x_i, x_{l,j}^1) f(x_{l,j}^*)$ \;
                    $\text{weights[i]} \pluseq w$ \;
                }
            }
        }
        $\hat{u}(x_i) = $ \text{sums[i] / weights[i]}  for all $i$ \;
    }
\end{algorithm}

In practice, finding the set of pairs of points where information can be passed $(\|x_i - x_l\| < r_l)$
can be performed efficiently with a bounding volume hierarchy of spheres.
Both approaches ensure the worst-case variance does not deteriorate when information is passed from nearby points and only require the start point, end point, and first step of each walk.

\section{Caching} \label{sec:caches}

We have only considered information reuse as post-processing for walks that have already been run.
The ability to reuse information also changes where we choose to begin our walks.
In this section, we show that our method enables estimates of $u$ on arbitrary subsets of $\Omega$ with walks starting on a grid.

First, we assume that there exists some $\delta > 0$ such that we are only interested in $u$ at least $\delta$ from the boundary.
We can trivially set $\delta = \varepsilon$ because the original Walk on Spheres algorithm terminates walks within $\varepsilon$ of the boundary,
although we will see that, when possible, larger choices of $\delta$ are advantageous.
For example, if we are only interested in the thermal profile of a sensitive component that is far from the boundary.
We can then construct a cache $C$ of size $O(\delta^{-d})$ to produce estimates at any point of interest.

\begin{lemma} \label{lem:cache}
Let $\delta > 0$ be arbitrary and $D_\delta = \{x \in \Omega: \operatorname{dist}(x, \partial \Omega) \ge \delta\}$ be the subset of $\Omega$ where we are interested in $u$.
There exists a cache set $C \subseteq D_\delta$ such that running walks from each element of $C$ yields estimates of $u(x)$ for all $x \in D$ using either \Cref{alg:eql} or \Cref{alg:var}.
\end{lemma}
\begin{proof}
We will prove this result under the looser condition of \Cref{alg:var}, but a similar construction holds for \Cref{alg:eql}.
Let $L$ be a $d$ dimensional grid with edge lengths $\delta/2$, such that $L$ is a $\delta/2$-covering of $D$.
Then it suffices to use $C = L \cap D_{\delta/2}$.

To prove this, let $x \in D_\delta$ be arbitrary. We will show that there exists some $c \in C$ such that $\|x-c\| < \operatorname{dist}(c, \partial \Omega)$, 
meaning walks from $c$ can be used to estimate $u(x)$.
There exists some $c \in L$ such that $\|x - c\| \le \delta/2$ because $L$ is a $\delta/2$-covering of $\Omega$.
This $c$ is also in $D_{\delta/2}$ because by the triangle inequality
\begin{align}
\operatorname{dist}(c, \partial \Omega) + \|x-c\| \ge \operatorname{dist}(x, \partial \Omega) \ge \delta \\
\implies \operatorname{dist}(c, \partial \Omega) \ge \delta / 2.
\end{align}
Therefore $c \in D_{\delta/2}$ from which we can conclude $\operatorname{dist}(c, \partial \Omega) \ge \delta / 2$ and $c \in C$, completing the proof.
\end{proof}

Importantly, the worst-case scaling significantly outperforms existing methods, with $O(\delta^{-d})$ compared to $O(m)$ size caches, as $\delta$ does not depend on $m$.
Treating the cache size as constant with respect to $m$ and $n$, this allows our algorithm have runtime $O(n) + O(mn)$ to run walks and pass information respectively.

To illustrate a case where our fixed size cache is helpful, let $\Omega = (-1,1)^2$ and $m$ be arbitrary.
Estimate $u$ at $m$ points in the subdomain $(-\frac{1}{2}, \frac{1}{2})^2$.
The cache construction of \Cref{lem:cache} with $\delta = \frac{1}{2}$ would return a cache of size $2^2 \left( \frac{\sqrt{8}}{0.5}\right)^2=128$.
In this case, we could do better than \Cref{lem:cache} by starting all walks from the origin and pass them to any point in $(-\frac{1}{2}, \frac{1}{2})^2$.
This allows us to estimate $u$ at any point in our subdomain arbitrarily well by simply increasing $n$.

\section{Numerical Experiments} \label{sec:experiments}

We demonstrate the performance of our algorithms on several example problems.
Comparisons between Walk on Spheres and mesh-based methods can be found in Sawhney and Crane \cite{sawhney2020monte} and consequently are not considered here.
In our example problems, we show a reduction in variance for a fixed number of walks and show performance scaling with problem size.
Although not the focus of this work, we discuss some ways in which performance could be further optimized, again noting the asymptotic improvement in \Cref{sec:caches}.
We test our new algorithms in two settings: post-processing and cache selection.

In the post-processing setting, our algorithms are as written in \Cref{alg:eql,alg:var}, using the tighter variance bound from our supplemental material.
We are given a domain $\Omega$, a boundary function $f$ on $\partial \Omega$, a number of walks $n$, and a set of points $x_1, \ldots, x_m \in \Omega$ where we estimate $u(x)$.
The same number of walks $n$ are performed from each starting point $x_1, \ldots, x_m$ via \Cref{alg:wos} and then the data from these walks are post-processed via \Cref{alg:eql,alg:var}.

In the cache selection setting, the information reuse scheme informs which points in $\Omega$ are used as starting points for the Walk on Spheres algorithm.
As discussed in \Cref{sec:caches}, we can observe an asymptotic speedup when the points $x_1, \ldots, x_m$ are sufficiently far from the boundary.

\begin{figure}
\centering
\includegraphics[scale=0.4]{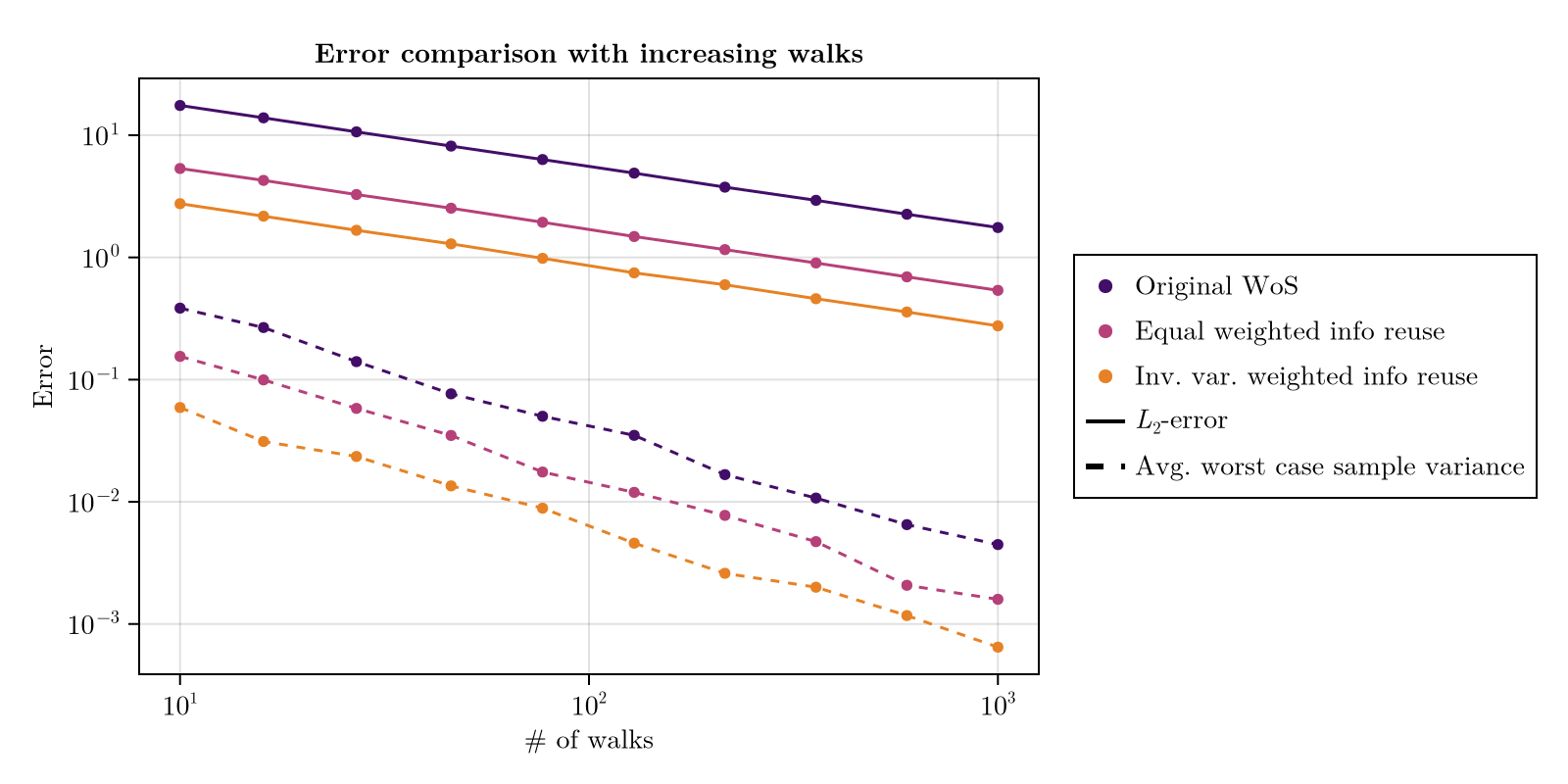}
\caption{Average performance of original Walk on Spheres when solving \Cref{eq:exp1} as the number of walks per point grows over 30 independent runs.}
\label{fig:exp1_asymptotic}
\end{figure}

\begin{figure*}[ht]
    \centering
    \begin{subfigure}{0.3\textwidth}
        \centering
        \includegraphics[scale=0.2]{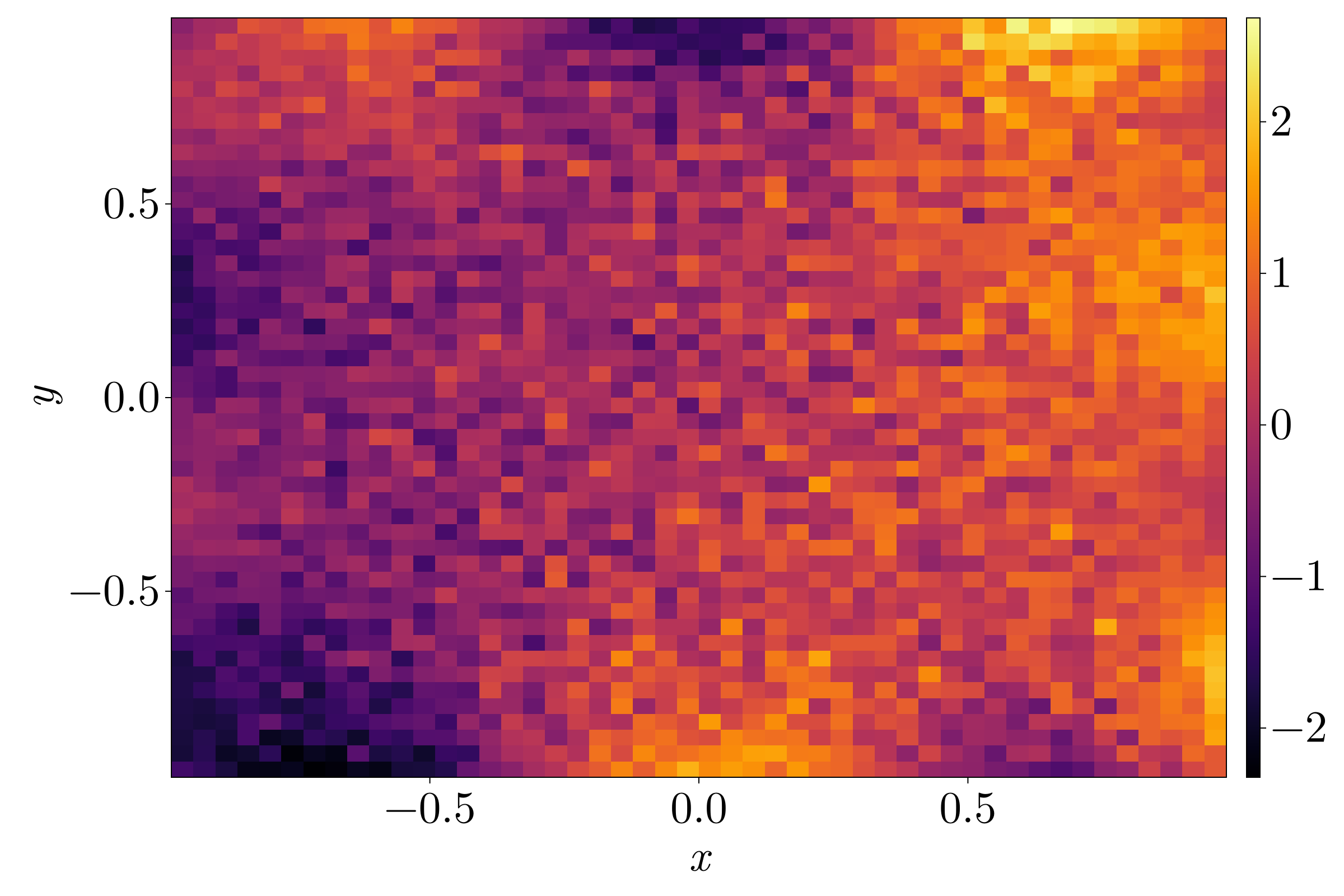}
        \caption{\centering \Cref{alg:wos}}
        \label{fig:taba}
    \end{subfigure}%
    \begin{subfigure}{0.3\textwidth}
        \centering
        \includegraphics[scale=0.2]{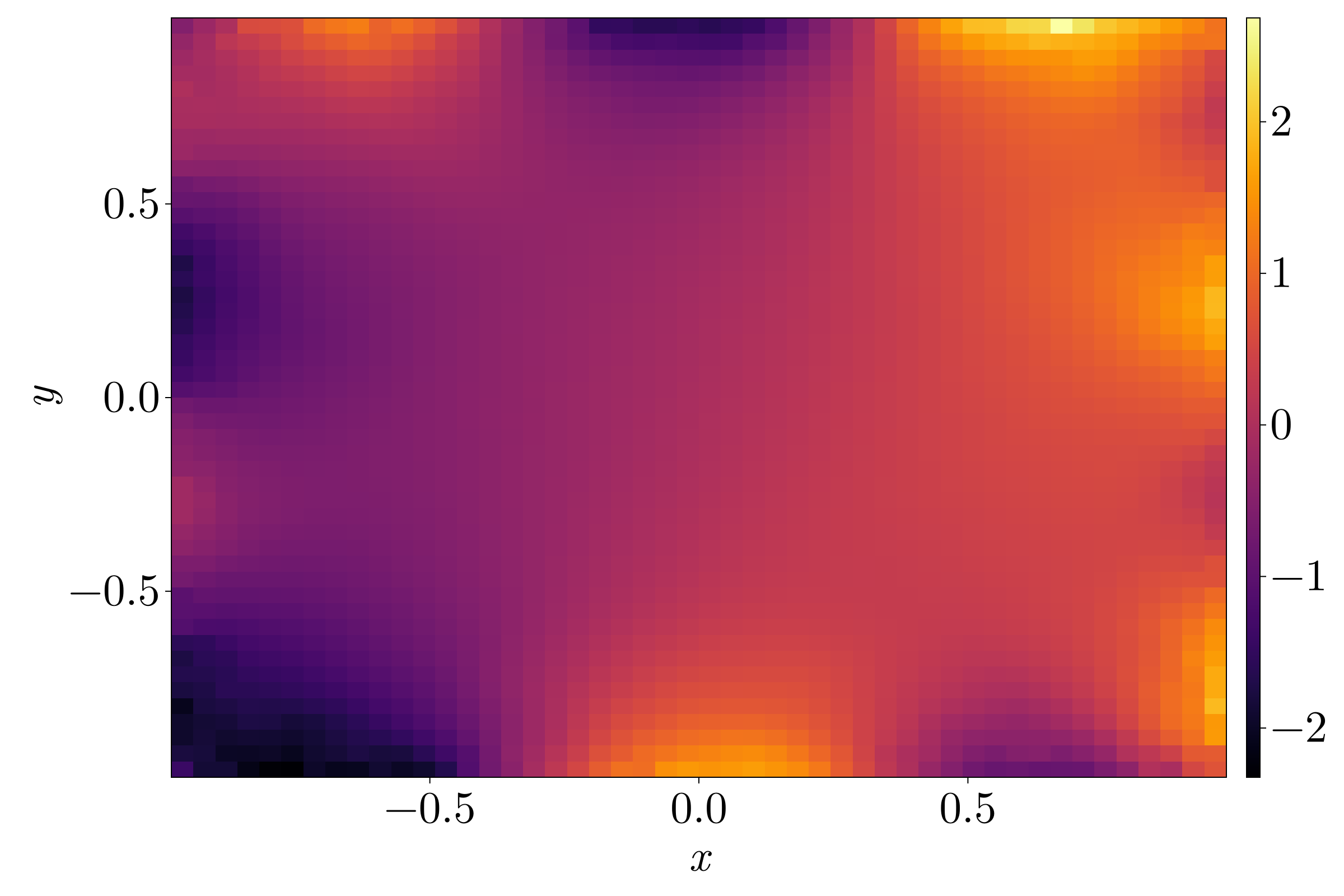}
        \caption{\centering \Cref{alg:eql} (ours)}
        \label{fig:tabb}
    \end{subfigure}%
    \begin{subfigure}{0.3\textwidth}
        \centering
        \includegraphics[scale=0.2]{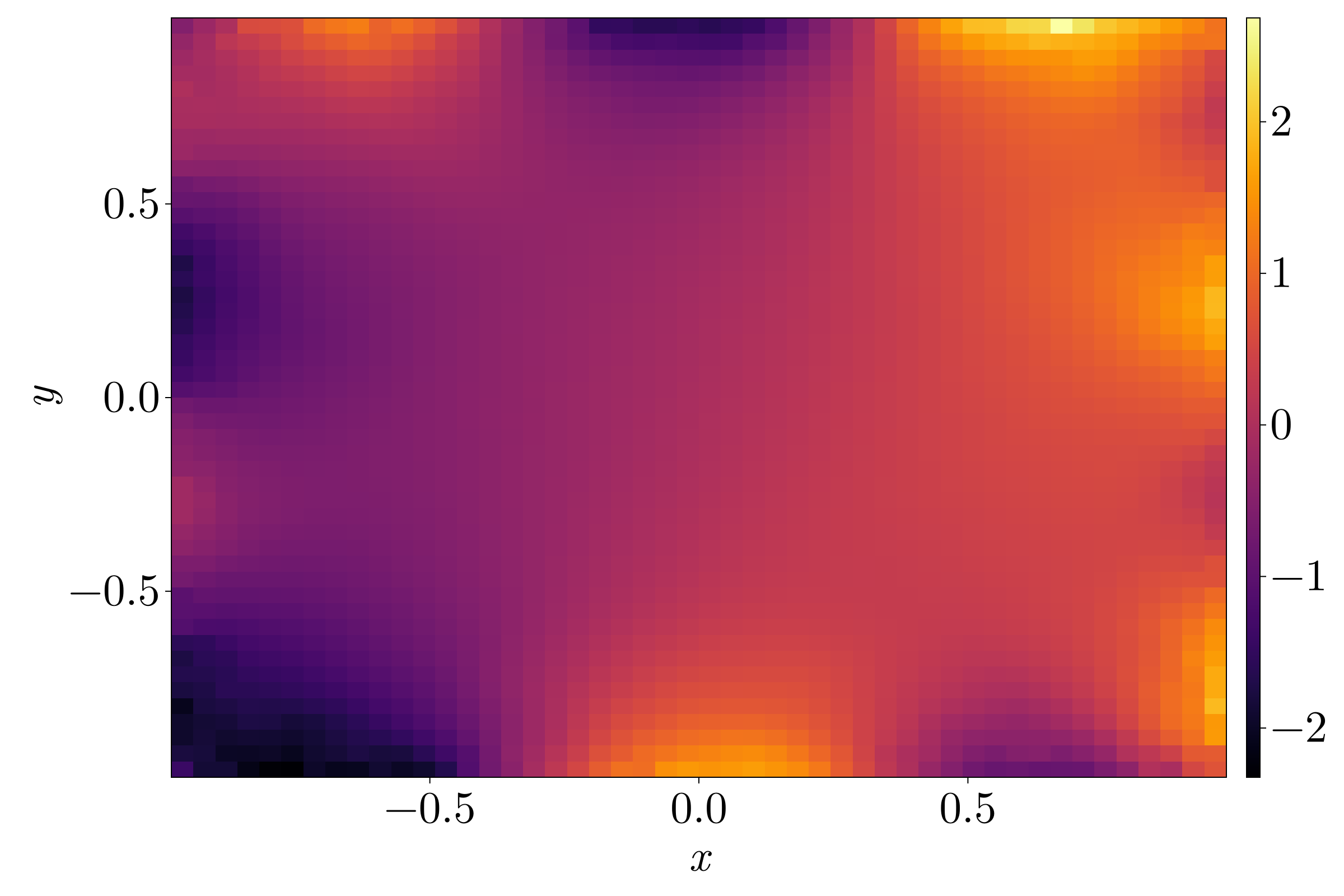}
        \caption{\centering \Cref{alg:var} (ours}
        \label{fig:tabc}
    \end{subfigure}
    \caption{Performance of \Cref{alg:wos,alg:eql,alg:var} on \Cref{eq:exp1} with 10 walks per point.}
    \label{fig:exp1_sol}
\end{figure*}

We demonstrate the performance in both settings on two example problems.
The first solves Laplace's equation on $[-1,1]^2$ with boundary function
\begin{align} \label{eq:exp1}
    f(x,y) = \begin{cases}
        -\sin(2\pi y) + x & x = -1\\
        \sin(2\pi y) + x  & x = 1\\
        -2\cos(\frac{3}{2}\pi x) + x & y=-1 \\
        2\cos(\frac{3}{2}\pi x) + x & y=1
    \end{cases}
\end{align}
on an evenly spaced $48 \times 48$ grid.
The solutions to this problem with 10 walks per point using each algorithm are shown in \Cref{fig:exp1_sol}.
We present the performance first with a fixed number of walks and then as the number of walks per point $n$ grows in \Cref{fig:exp1_asymptotic}.

The original Walk on Spheres algorithm (\Cref{alg:wos}) estimates the solution independently across points, producing a salt-and-pepper appearance.
The center of the solution is well resolved by the information reuse schemes (\Cref{alg:eql,alg:var}) while the solution quality near the boundary remains similar.
This is a function of the geometry.
Spheres centered near the center of the domain have larger radii and can pass their walk information to more points, while the opposite is true near the boundary.
We also observe similar results in \Cref{fig:teaser}, where we solve Laplace's equation in a 3d box containing the University of Wisconsin-Madison's mascot Bucky with periodic boundary conditions on the box and color-dependent boundary conditions on the mesh of Bucky.

The asymptotic results for this example are presented in \Cref{tab:bucky}.
We vary the number of walks from each point for each algorithm and present the total runtime, runtime of post-processing for variance reduction, and both the average and maximum variance of our estimates over the solution grid. 
We obtain orders of magnitude improvement in the average variance at a runtime cost of less than $1\%$.

\begin{table}[h]
    \centering
    \small
    \caption{Runtime and Variance Scaling for \Cref{fig:teaser} on an Intel(R) Xeon(R) CPU E5-2620 v3 @ 2.40GHz CPU with 32GB of memory using 12 threads} 
    \label{tab:bucky}
    \begin{tabular}{||c c c c c c||} 
     \hline
    Walks & Alg. & Total (s)& Post-process (s) & Avg. Var. & Max. Var. \\ [0.5ex] 
     \hline\hline
    5 & 0 & 9076 & 0 & 0.0684 & 0.24 \\ 
     \hline
    5 & 1 & 9069 & 1.73 &  0.010 & 0.24\\
     \hline
    5 & 2 & 9162 & 32.41 &  0.000284 & 0.24\\
     \hline
    10 & 0 & 18027 & 0 & 0.03 & 0.111\\
     \hline
    10 & 1 & 17978 & 2.29 & 0.00517 & 0.111 \\
    \hline
    10 & 2 & 18613& 57.7 & 0.000147 & 0.0933\\ [1ex] 
     \hline
    \end{tabular}
\end{table}

We show the asymptotic error in solving \Cref{eq:exp1} on an evenly spaced $48 \times 48$ grid in \Cref{fig:exp1_asymptotic}, measured both as the worst-case variance in our pointwise estimators and the $L^2$ error of the solution on the grid.
The $L^2$ error is vastly reduced for both information reuse schemes.
The worst-case variance shows that our information reuse schemes satisfy the bounds of \Cref{sec:var_reduce} as the worst observed variance across solution points is bounded from above by the worst variance observed by the original Walk on Spheres algorithm.
By both measures, our methods outperform the original Walk on Spheres algorithm.

We also stress the advantage of an arbitrary cache choice.
In \Cref{sec:caches}, we gave a general method for constructing a cache where walks are to be started.
Provided points where we estimate $u$ are away from the boundary, we can improve the runtime as the number of points $m$ grows.
To demonstrate this, we again solve \Cref{eq:exp1}, but only aim to estimate $u$ on $[-0.5, 0.5]^2$ on an increasingly dense square grid while keeping the cache constant.
We fix the cache as an evenly spaced $5\times 5$ grid on $[-0.5, 0.5]^2$ and only start walks at these $25$ points.
The original Walk on Spheres algorithm runs $n$ walks from each point, and we run an equivalent number of walks distributed evenly over our cache.
Therefore, all algorithms perform the same number of walks.
\Cref{fig:exp1_asymptotic} shows the $L^2$ error of our solution as the size of the grid increases for both 10 and 100 walks per point.

\begin{figure}
\centering
\includegraphics[scale=0.4]{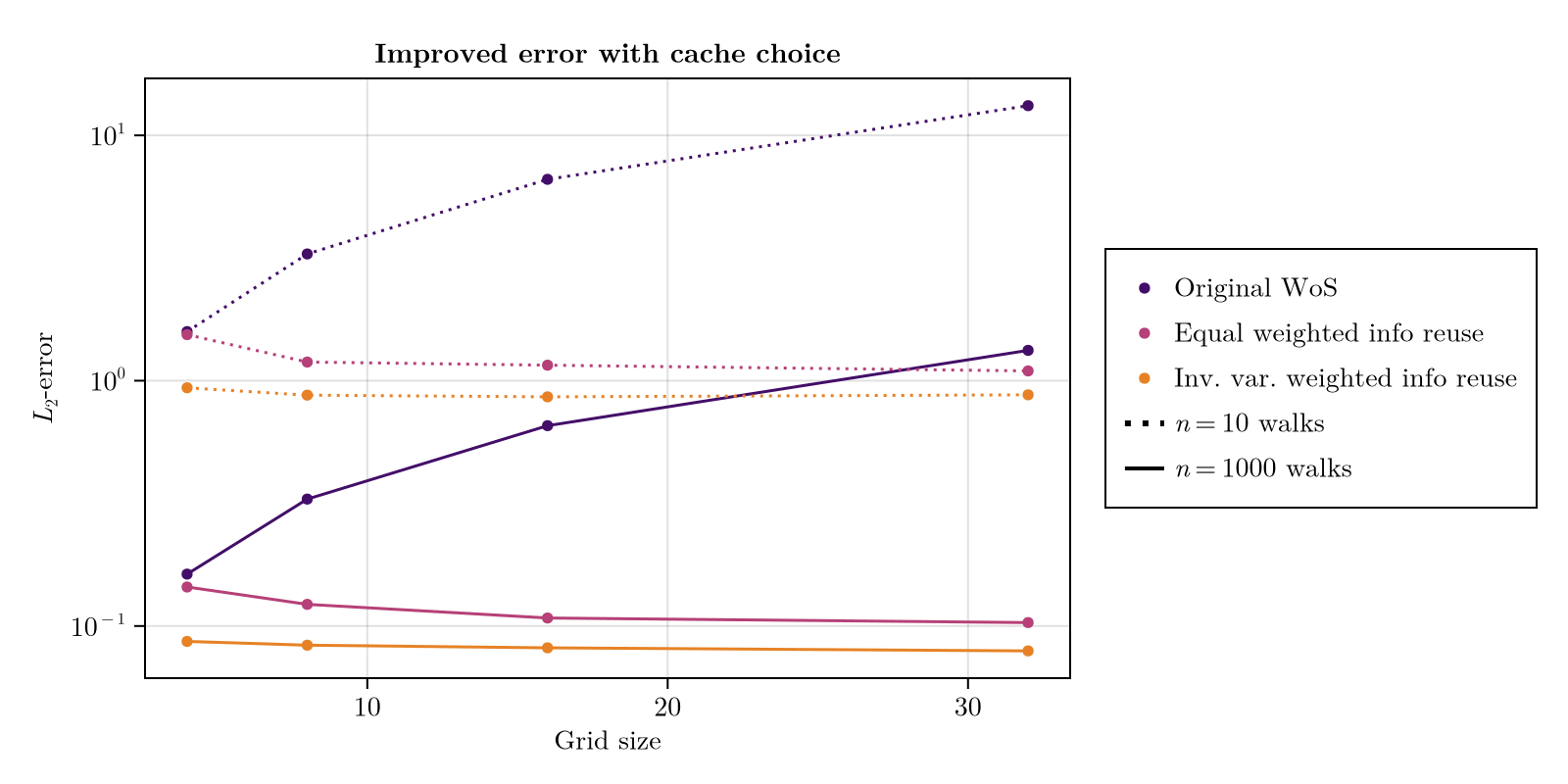}
\caption{Average performance of original Walk on Spheres when solving \Cref{eq:exp1} with 10 and 1000 walks per point over 30 runs.
Information reuse schemes use a fixed 25 element cache and run the same number of total walks as original Walk on Spheres.}
\label{fig:exp2_asymptotic}
\end{figure}

We see that as the grid becomes more dense, the original Walk on Spheres $L^2$ error increases (note the vector of errors also grows in length).
However, the $L^2$ errors for our information reuse algorithms remain roughly the same, because as the number of grid points grows, the number of unbiased estimators at each grid point grows as well due to information reuse.

To demonstrate the performance of our algorithm on complex, non-smooth geometry, we present an experiment where the domain boundary is inspired by the boundary of a fractal Julia set \cite{barnsley1988science}. 

\begin{figure*}[ht]
    \centering
    \begin{subfigure}{0.3\textwidth}
        \centering
        \includegraphics[scale=0.07]{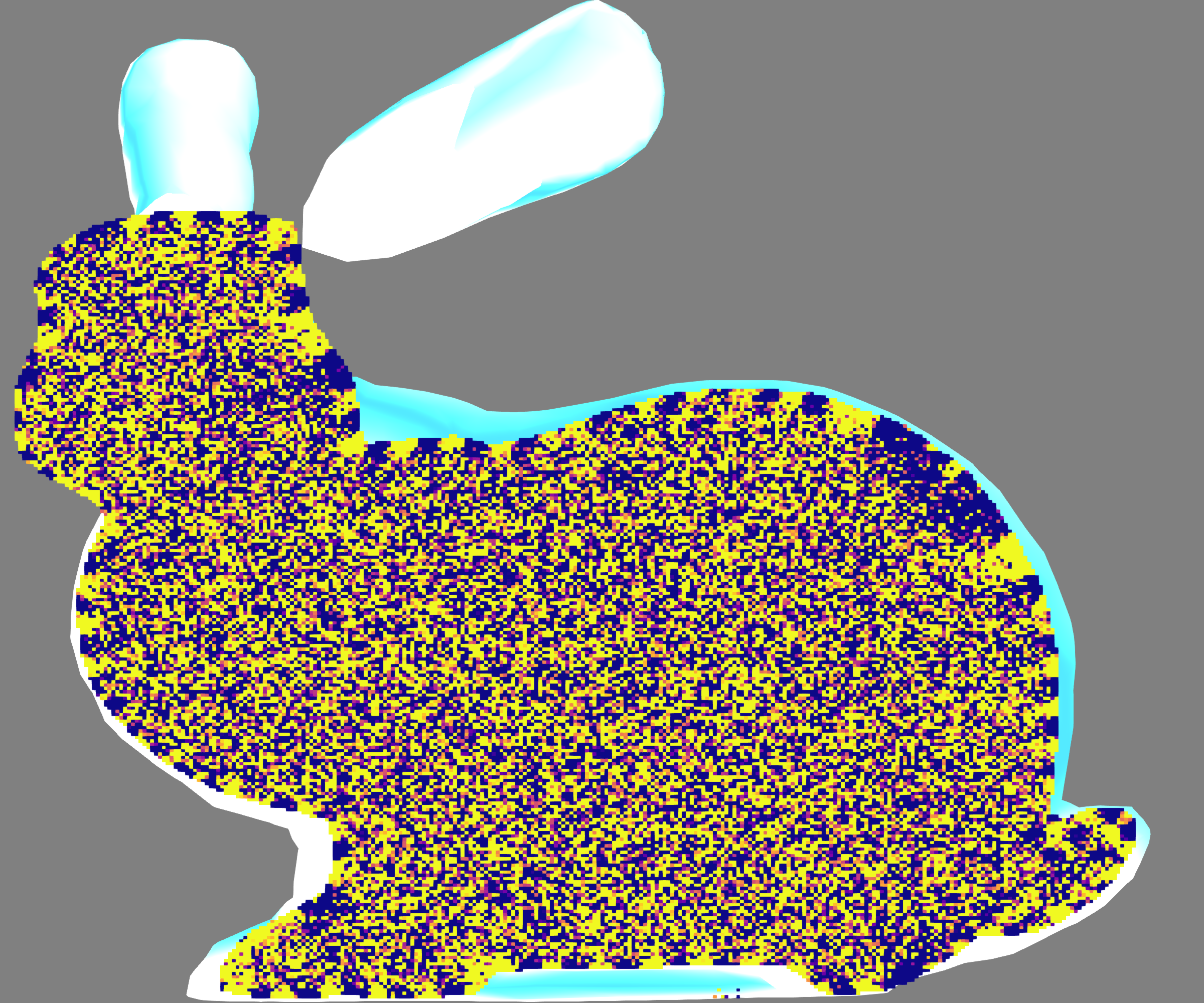}
        \caption{\centering \Cref{alg:wos}}
        \label{fig:taba}
    \end{subfigure}%
    \begin{subfigure}{0.3\textwidth}
        \centering
        \includegraphics[scale=0.07]{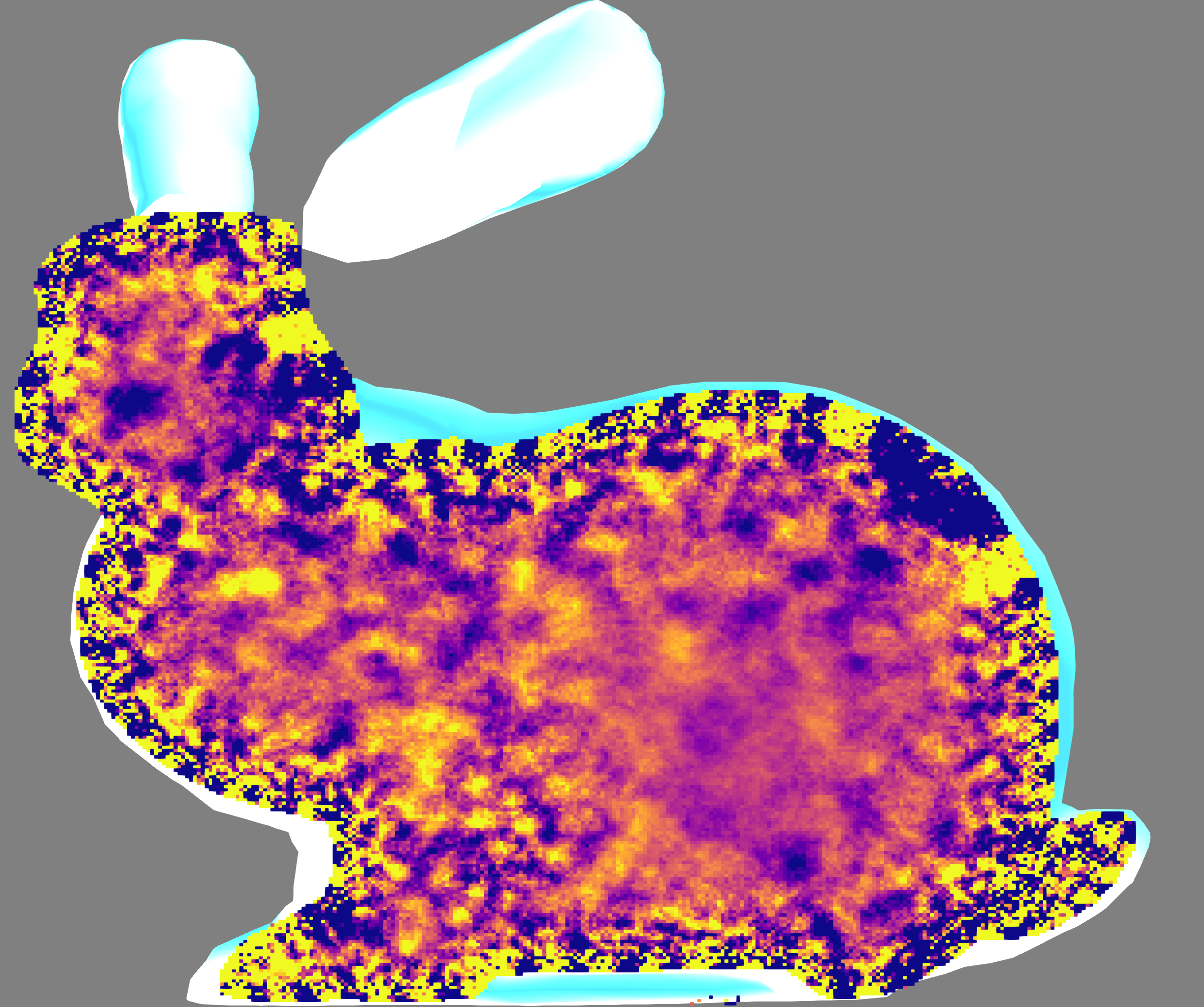}
        \caption{\centering \Cref{alg:eql}}
        \label{fig:tabb}
    \end{subfigure}%
    \begin{subfigure}{0.3\textwidth}
        \centering
        \includegraphics[scale=0.07]{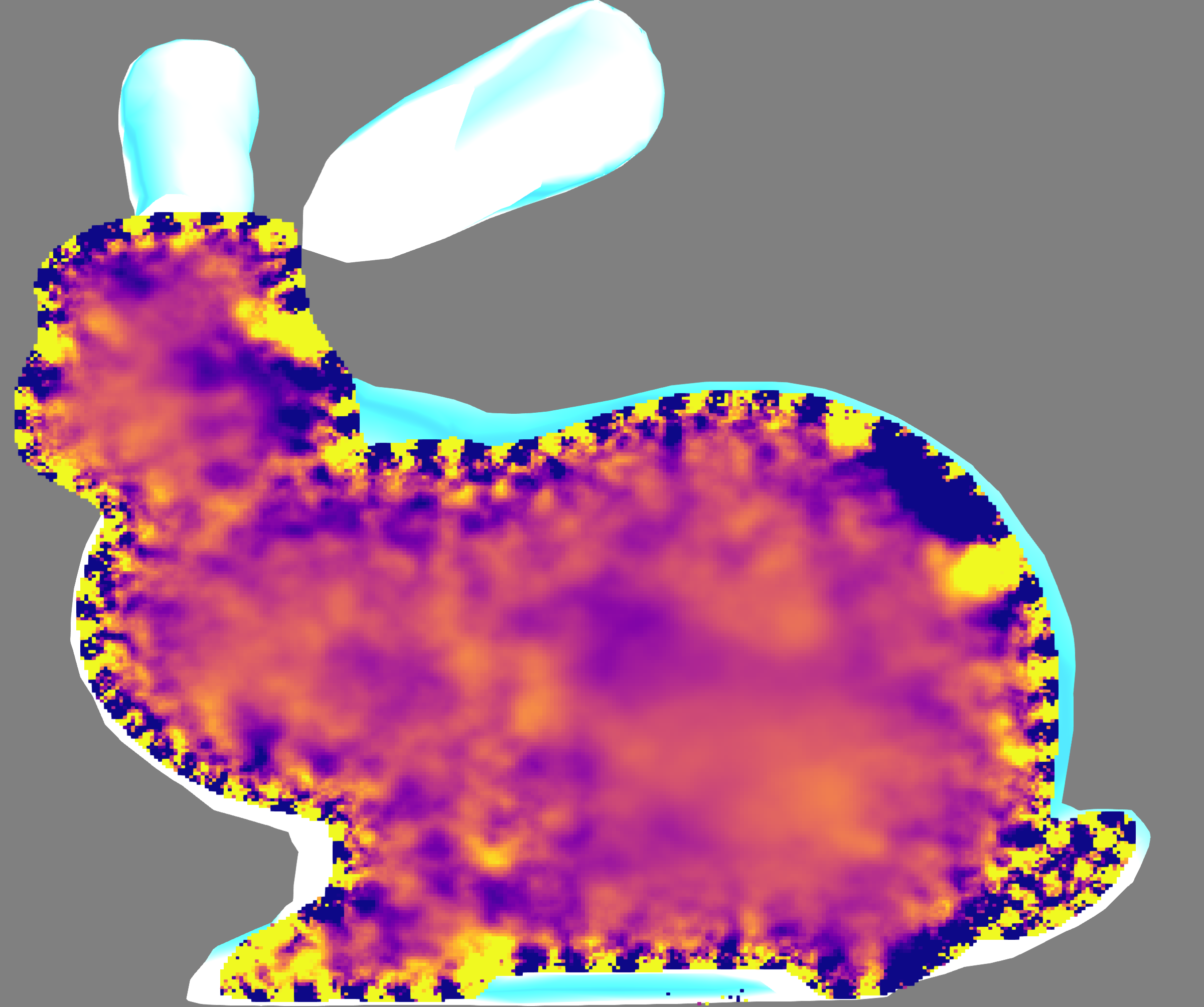}
        \caption{\centering \Cref{alg:var}}
        \label{fig:tabc}
    \end{subfigure}
    \caption{Performance of \Cref{alg:wos,alg:eql,alg:var} on the Stanford bunny with 10 walks per point.}
    \label{fig:bunny}
\end{figure*}

\begin{figure*}[ht]
    \centering
    \begin{subfigure}{0.3\textwidth}
        \centering
        \includegraphics[scale=0.07]{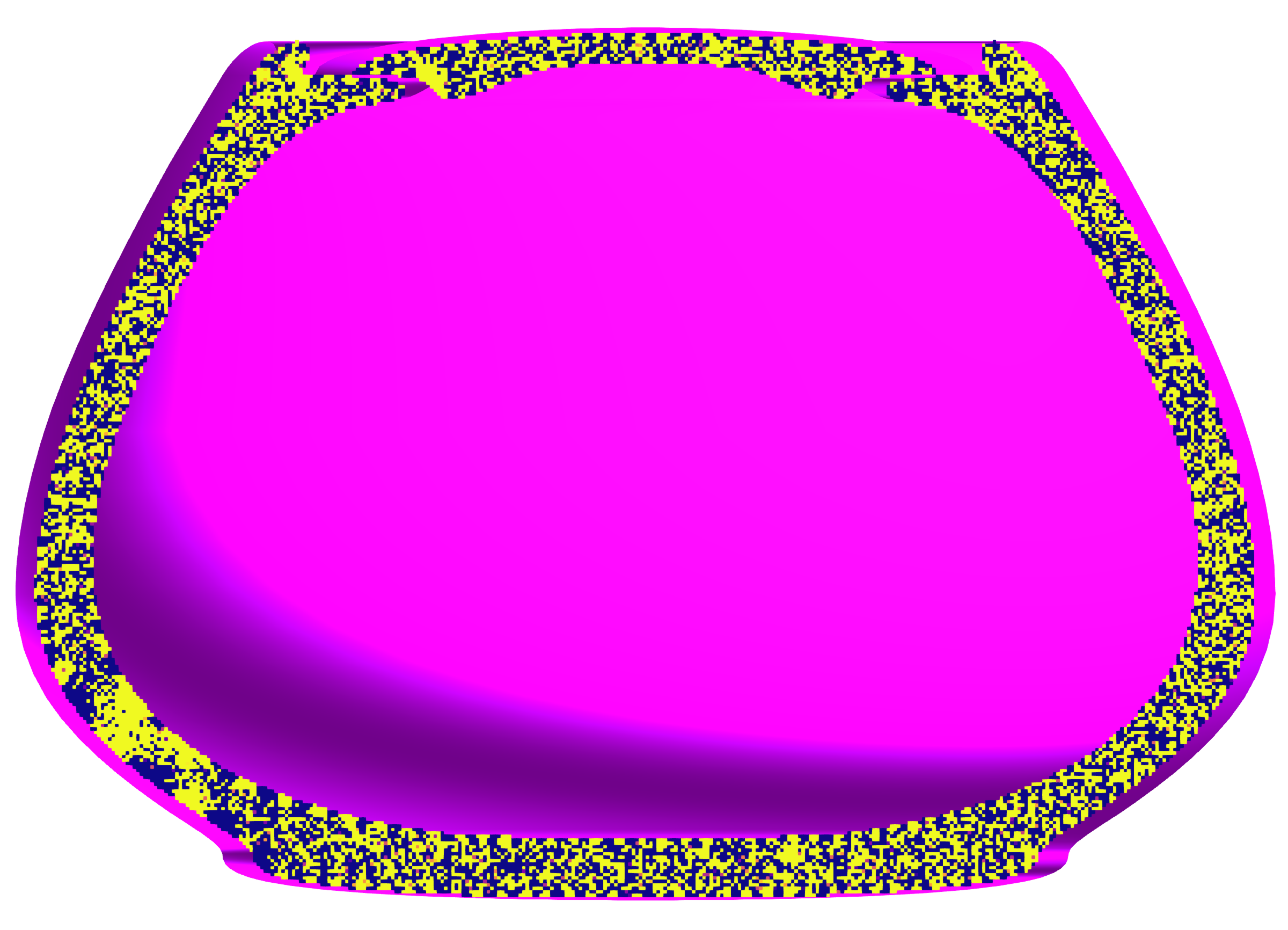}
        \caption{\centering \Cref{alg:wos}}
        \label{fig:taba}
    \end{subfigure}%
    \begin{subfigure}{0.3\textwidth}
        \centering
        \includegraphics[scale=0.07]{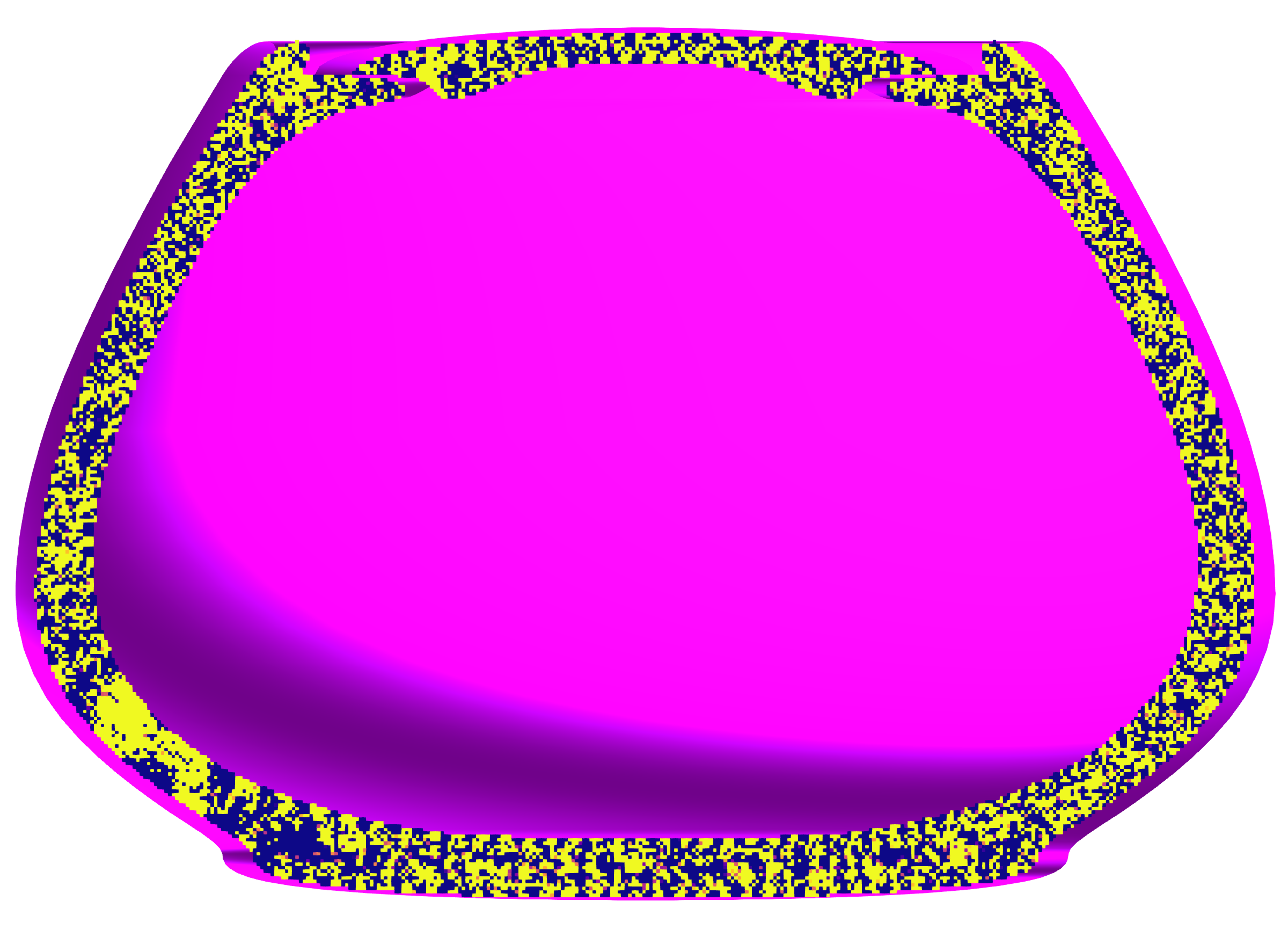}
        \caption{\centering \Cref{alg:eql}}
        \label{fig:tabb}
    \end{subfigure}%
    \begin{subfigure}{0.3\textwidth}
        \centering
        \includegraphics[scale=0.07]{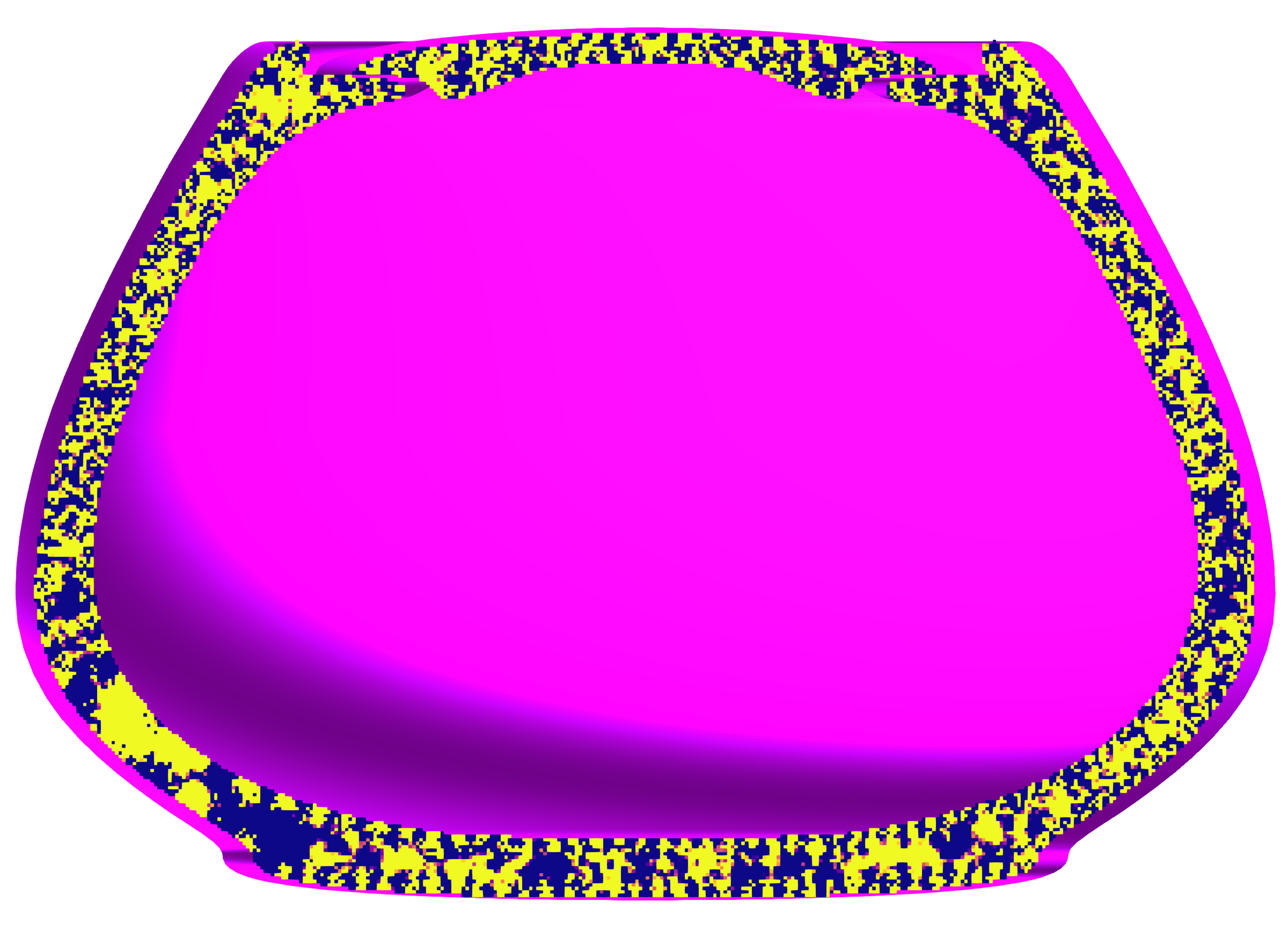}
        \caption{\centering \Cref{alg:var}}
        \label{fig:tabc}
    \end{subfigure}
    \caption{Performance of \Cref{alg:wos,alg:eql,alg:var} on the Utah teapot interior  with 10 walks per point.}
    \label{fig:teapot}
\end{figure*}

We also present example results on the Stanford Bunny and Utah Teapot in \Cref{fig:bunny,fig:teapot}
Both solutions are slightly under-resolved to show the spatial structure that our algorithms exploit.
This results in a smoother approximation of the solution even with few walks.
In both cases we obtain at least an order of magnitude improvement in the average variance at marginal runtime cost (see \Cref{tab:bucky,tab:teapot}).

\begin{table}[h]
    \centering
    \small
    \caption{Runtime and Variance Scaling for \Cref{fig:bunny} on an Intel(R) Xeon(R) CPU E5-2620 v3 @ 2.40GHz CPU with 32GB of memory using 12 threads} 
    \label{tab:bunny}
    \begin{tabular}{||c c c c c c||} 
     \hline
    Walks & Alg. & Total (s)& Post-process (s) & Avg. Var. & Max. Var. \\ [0.5ex] 
     \hline\hline
    4 & 0 & 1142 & 0 & 24.07 & 118.5 \\ 
     \hline
    4 & 1 & 1177 & 1.25 &  3.88 & 116.9\\
     \hline
    4 & 2 & 1141 & 25.8 &  0.097 & 72.9\\
     \hline
    10 & 0 & 2910 & 0 & 9.65 & 31.6\\
     \hline
    10 & 1 & 2857 & 1.80 & 1.55 & 29.4 \\
    \hline
    10 & 2 & 2890& 49.48 & 0.043 & 13.6\\ [1ex] 
     \hline
    \end{tabular}
\end{table}

\begin{table}[h]
    \centering
    \small
    \caption{Runtime and Variance Scaling for \Cref{fig:teapot} on an Intel(R) Xeon(R) CPU E5-2620 v3 @ 2.40GHz CPU with 32GB of memory using 12 threads} 
    \label{tab:teapot}
    \begin{tabular}{||c c c c c c||} 
     \hline
    Walks & Alg. & Total (s)& Post-process (s) & Avg. Var. & Max. Var. \\ [0.5ex] 
     \hline\hline
    4 & 0 & 288.7 & 0 & 1153.9 & 3288.1\\ 
     \hline
    4 & 1 & 291.2 & 0.03 &  923.3 & 3166.4\\
     \hline
    4 & 2 & 285.7 & 0.10 &  49.0 & 2225.4\\
     \hline
    10 & 0 & 635.6 & 0 & 462.0 & 988.5\\
     \hline
    10 & 1 & 616.4 & 0.02 & 367.2 & 933.0 \\
    \hline
    10 & 2 & 594.3 & 0.11 & 20.9 & 692.8\\ [1ex] 
     \hline
    \end{tabular}
\end{table}

This performance can be further accelerated by leveraging GPUs and the independence of the walks, which is the subject of on-going work.
\begin{figure}[H]
    \centering
    \includegraphics[scale=.15]{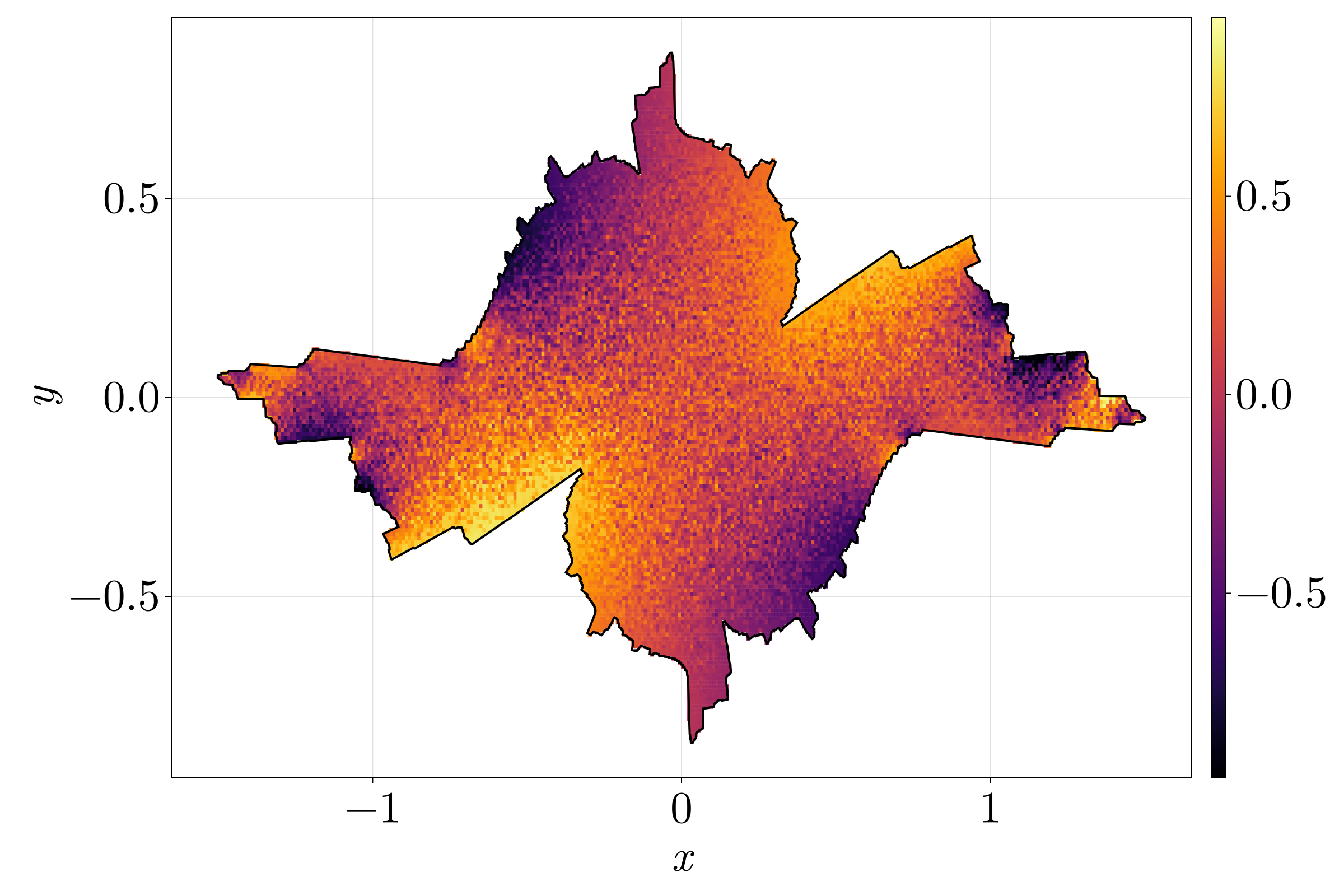}%
    \includegraphics[scale=.15]{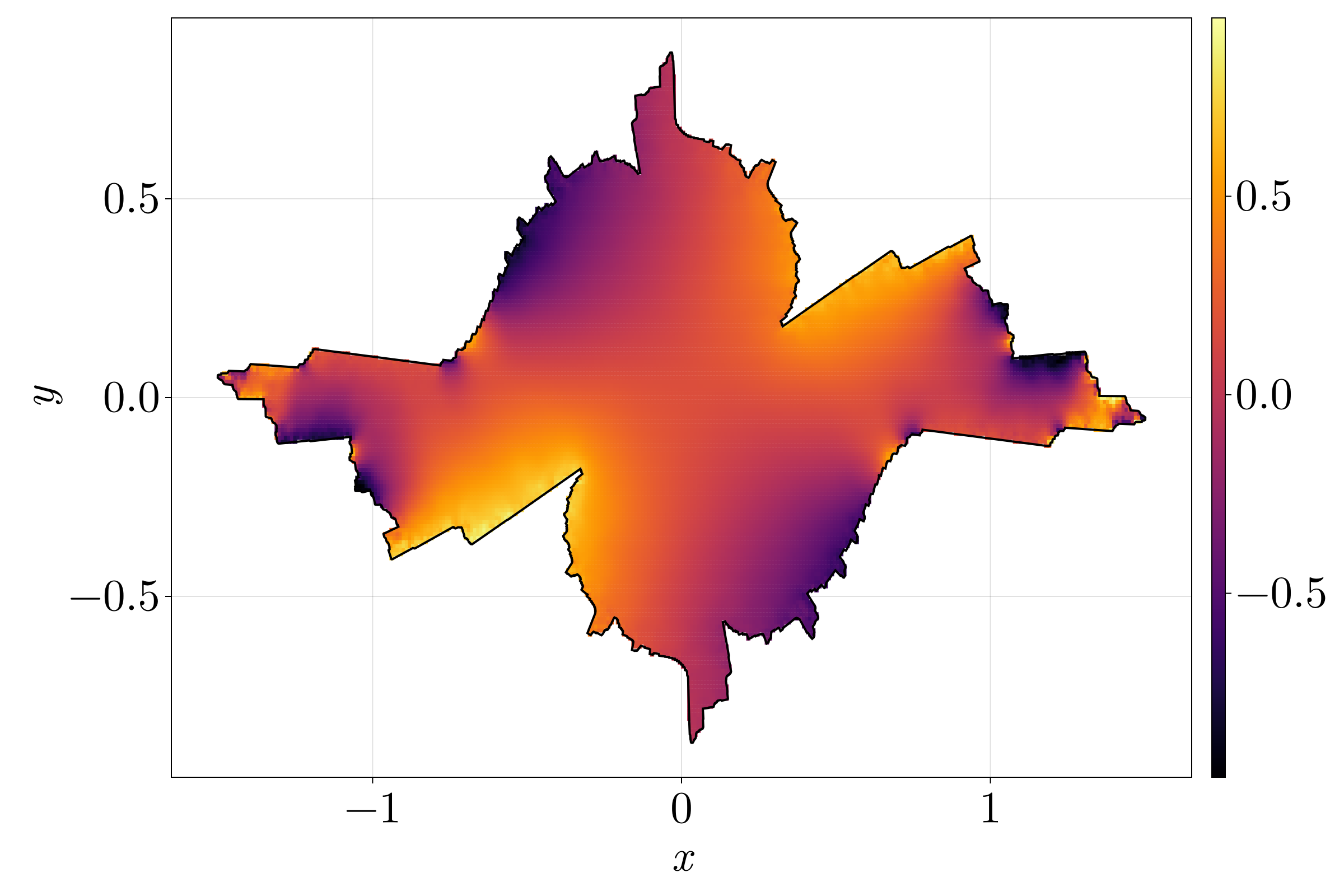}
    \caption{Solution to the Laplace problem on the Julia set inspired boundary using 8 walks per point without variance reduction (left) and with variance-weighted information reuse (right).}
    \label{fig:julia_set}
\end{figure}

\section{Discussion and conclusions}

The methods and bounds we describe improve the asymptotic runtime of Walk on Spheres with caching and allow for the arbitrary choice of cache points.
This produces better estimates of solutions to Laplace's equations without the use of a mesh and opens the door to using these methods in more complicated geometries with fewer computational resources.

The methods we propose in provably reduce the variance of our estimator and provide a novel method for information reuse in the Walk on Spheres algorithm.
We also show how to use an arbitrary cache.
Provided that we do not try to estimate $u$ arbitrarily close to the boundary, we provide a cache construction to obtain an asymptotic runtime improvement.
The improved variance of our proposed algorithm is also demonstrated.
Moreover, the variance bounds provide new avenues to improve the performance of the Walk on Spheres algorithm.

For example, our information reuse scheme allows for the use of an arbitrarily chosen cache.
This could be an adaptively sampled cache, a cache constructed using prior knowledge of the problem, or the construction in \Cref{lem:cache}. 
The size of the cache constructed in \Cref{lem:cache} demonstrates the gains that can be made by taking advantage of the flexibility in cache choice afforded by our information reuse.
A smaller cache than the one in \Cref{lem:cache} is certainly possible.
For example, $\delta$ in \Cref{lem:cache} could be measured once given a set of points of interest.
Notably, the cache size does not increase with the density of points at which we would like solutions; it only depends on how far those points are from the boundary.
This is particularly useful when Walk on Spheres is used to estimate $u$ on some subset of $\Omega$ without performing a global solve.

Just as we could construct better caches, it is also possible to improve the bounds in \Cref{sec:var_reduce}.
For example, tighter bounds could potentially be derived from assumptions on the regularity of $f$.
Popoviciu's inequality is rather blunt in this setting, and although it is possible to construct problems where these bounds are tight, they are quite loose for most problems in practice.
Even with the existing bounds, our information passing scheme could be extended to other quantities of interest such as gradients, Hessians, or the solutions of other classes of partial differential equations.
The reweighting detailed here is viable for more general PDEs, although the bounds would need to be revised.

\begin{acks}
This work was supported by Department of Energy Award Nos. DE-SC0024548, DE-FG02-93ER54222, DE-AC52-07NA27344 (HiFiStell), a grant from the Simons Foundation (No. 560651, D B), and a Department of Energy Summer Undergraduate Research Internship (SULI) program. This research used resources of the National Energy Research Scientific Computing Center (NERSC), a Department of Energy Office of Science User Facility using NERSC award FES-ERCAP0032166.
\end{acks}

\bibliographystyle{ACM-Reference-Format}
\bibliography{wos}

\end{document}